\documentclass{JHEP3}
\long\def\symbolfootnote[#1]#2{\begingroup%
\def\thefootnote{\fnsymbol{footnote}}\footnote[#1]{#2}\endgroup} 

\usepackage{epsfig}

\newcommand{\bea}{\begin{eqnarray}}
\newcommand{\eea}{\end{eqnarray}}

\newcommand{\la}{\langle}
\newcommand{\ra}{\rangle}
\newcommand{\be}{\begin{equation}}
\newcommand{\ee}{\end{equation}}
\newcommand{\tr}{{\rm Tr}}
\newcommand{\call}{{\cal C}}
\newcommand{\area}{{\cal A}}
\newcommand{\vpot}{{\mathit a}}
\newcommand{\LL}{{\cal L}}

\author{R. Narayanan
\\Department of Physics, Florida International University, Miami,
FL 33199, USA\\E-mail: \email{rajamani.narayanan@fiu.edu}}
\author{ H. Neuberger
\\ Rutgers University, Department of Physics and Astronomy,
Piscataway, NJ 08855, USA\\E-mail: \email
{neuberg@physics.rutgers.edu} }

\title{Two dimensional fermions in four dimensional YM}

\abstract {Dirac fermions in the fundamental representation of $SU(N)$ 
live on a two dimensional torus flatly embedded in $R^4$.
They interact with a four dimensional $SU(N)$ Yang Mills vector potential preserving
a global chiral symmetry at finite $N$. As the size of the torus in units of $\frac{1}{\Lambda_{SU(N)}}$ 
is varied from small to large, 
the chiral symmetry gets spontaneously broken in the
infinite $N$ limit.}


\preprint{}

\begin{document}

\section{Introduction}

When the local description of a physical system includes strong non-linearities, one tries to trade the original variables 
for ``better'' variables, in terms of which nonlinear effects become weaker. A goal more likely within reach is to 
find ``good'' observables whose behavior 
displays the essence of the nonlinearity. 
In the context of relativistic field theory these 
observables are non-local in terms of the original variables. As such, their definition requires 
specific study of their renormalization. 

In this paper we introduce a non-local operator in four dimensional gauge theory which 
generalizes the Wilson loop associated with a closed one dimensional curve in spacetime 
to an object associated with a closed two dimensional surface embedded in spacetime.

\section{Wilson loop operators}

\subsection{General properties}

The nonlinear dynamics of pure $SU(N)$ Yang Mills gauge theory on Euclidean $R^4$ is
best captured by nonlocal observables. The Wilson loop operators are standard.
\be
\tr_r \la  {\cal P} e^{i\oint_\call A\cdot dx}\ra \equiv \tr_r \la  \Omega(\call)\ra =W_r(\call)
\ee
Here $r$ denotes an irreducible representation of $SU(N)$, ${\cal P}$ denotes
path ordering round a closed, smooth, non-selfintersecting curve $\call$, and $A$ is
the hermitian Yang Mills connection of $SU(N)$ Yang Mills
theory in $R^4$ with ordinary action. The $\theta$ parameter is set to zero. 

The number of boxes modulo $N$ in the Young pattern describing $r$ is 
the $N$-ality $k$. Assume that 
the curve $\call$ has a unique minimal spanning area $\area$
in the standard metric of $R^4$. A major feature
of the theory is confinement: For $k\ne 0$ one has 
\be
\log [W_r(\call)] {\mathop {\sim}_{\area\to\infty}} -\sigma_k \area
\label{conf}
\ee
Here the limit $\area\to\infty$ is taken at fixed 
loop shape, $\hat\call$, by uniformly dilating the loop until the required
value of $\area$ is achieved. $\sigma_k >0 $, the $k$-th string tension, 
does not depend on $\hat\call$. Since
we have set the $\theta$-parameter to zero, charge conjugation gives 
$\sigma_k=\sigma_{N-k}$. Thus, Wilson loop operators 
provide a precise definition of the property of confinement, quantitatively expressed by the $\sigma_k$. 

$\call$ can be parameterized by $\area$ and ${\hat\call}$.
In perturbation theory a dependence on $\area$ comes in only through the quantum
violation of classical scale invariance. 
The behavior of $W_r(\call)$ as $\area\to 0$ is determined by perturbation theory
in terms of the scale $\Lambda_{SU(N)}$ which sets a convention for a 
running coupling constant  
at short scales, common to all observables. The perturbative expansion
of $W_r(\call )$ is defined only after making some choices to fix free parameters 
one encounters in the process of renormalization. These choices may 
extend beyond perturbation theory, but should not affect $\sigma_k$.

The renormalization of $W_r (\call )$ is well
understood~\cite{gervaisneveu, dotsenkovergeles}. 
In perturbation theory one needs to expand $\log [ W_r (\call ) ]$ and
the needed ``exponentiation'' Feynman rules are known~\cite{expon}. 
The conclusion is that for every $r$ one has one 
operator dependent linear perimeter divergence and
all the other divergences are the same as in pure gauge theory. 
Imposing symmetry under conjugation of the renormalized Wilson loops 
leaves one new arbitrary real finite part for distinct 
self-conjugate representations, $r\cong {\bar r}$, or pairs
of distinct conjugate representations, $r\not\ncong {\bar r}$. 

\subsection{Fermionic representation}

We shall focus on the set of all
single column (totally antisymmetric) representations. This set
has then $\left [ \frac{N}{2}\right ]$ free real parameters.
These parameters can be counted also by simultaneously dealing with all the 
$W_r (\call )$, assembled into the average characteristic polynomial:
\be
P(z,\call )=\la \det [z+\Omega (\call ) ]\ra = \la \det [1+z\Omega^\dagger (\call ) ]\ra
\ee

The diagrammatic expansion rules for $\log P(z,\call )$ are obtained from a fermionic
representation. Let $X_1,X_2...X_n$ be $n$ arbitrary $N\times N$ matrices and
let $X=X_1X_2...X_n$ be their matrix product. We then have the following 
identity~\cite{three-d, iden}:
\be
\int \prod_{j=1}^n [d\bar\psi_j d\psi_j ] e^{\sum_{j=1}^n [
\bar\psi_j X_j\psi_{j+1} -\bar\psi_j \psi_j 
]} = \det(1+X)
\ee
Here $\psi_{n+1}\equiv -\psi_1$ and each $\psi_j$ has $N$ components. 
Taking a formal continuum limit we get~\cite{oddoverlap}:
\be
\det[1+z\Omega^\dagger (\call ) ]=\int [d\bar\psi d\psi ] e^{\int_0^l d\sigma
\bar\psi(\sigma)[\partial_\sigma - \mu - i\vpot(\sigma)]\psi(\sigma)}
\ee
Here, 
\be
z=e^{-\mu l}
\ee
$\sigma$ parametrizes $\call$ by $x(\sigma)$. The parametrization 
$[\partial_\sigma x_\mu (\sigma )]^2=1$ identifies $l$ as the length of the loop.
The continuum 
one dimensional fermions $\bar\psi(\sigma),\psi(\sigma)$ obey anti-periodic
boundary conditions round the curve. The one dimensional vector potential 
$\vpot$ is defined by the four dimensional one:
\be
\vpot (\sigma) = A_\mu (x(\sigma ))
\frac {\partial x_\mu(\sigma)}{d\sigma}
\ee

The diagrams are read off from the following path integral:
\be
P(z,\call)=\int [dA_\mu ] [d\bar\psi d\psi ]  
\frac {e^{-\frac{1}{4g^2}\int d^4 x \tr F^2 +\int_0^l d\sigma
\bar\psi(\sigma)[\partial_\sigma - \mu - i\vpot(\sigma)]\psi(\sigma)}}{
\int [dA_\mu ] e^{-\frac{1}{4g^2}\int d^4 x \tr F^2}}
\ee
$g$ is the gauge coupling and $F_{\mu\nu}$ is the field strength made out of $A_\mu$.
One needs to choose a gauge fixing convention and add the needed ghosts.
This is done without reference to the fermions. 

The Feynman diagram expansion for $\log[P(z,\call)]$ consists of connected diagrams
which look exactly like those for the fermion contribution
to the free energy of ordinary QCD. Only the mathematical 
expressions for the $\bar\psi\psi A$ vertex and for the $\psi\bar\psi$
propagator differ.

If the diagrams are viewed in Euclidean configuration space the expression for the $\bar\psi\psi A$ vertex is simple. The fermion 
propagator, $S(\sigma)$, requires the inversion of the kernel taking
into account the boundary conditions. With $|\sigma|<l$ ($\sigma$ 
stands for $\sigma_1-\sigma_2$ with $0<\sigma_j<l$) 
it is given by~\cite{dunne}:
\be
S(\sigma)=e^{-\mu\sigma}\left [ \theta (\sigma) - \frac{1}{1+e^{\mu l}}\right ]
\ee
$S(\sigma)$ is diagonal in its suppressed $SU(N)$ indices. 
The dependence on $\call$ comes from connected vacuum diagrams
containing at least one fermion loop and makes a contribution
of at most order $N$ to the free energy if $g^2 N$ is kept fixed. 

$W_r$ for $r=N$ (the fundamental representation) can be extracted by taking $\mu\to\infty$
and picking out the term linear in $z$. For a given fermion loop the product
of the prefactors of $S(\sigma)$ is one. All propagators are now  $\theta$-functions  except one, which is $-z$. The fermion loop gives another overall $-1$. 
The dropped $\theta$ function is equal to 
zero if all other $\theta$ functions are equal to one. This recovers the 
pure gauge Feynman rules for the fundamental Wilson loop operator. To the same end, one
could have also used the limit $\mu\to -\infty$ ($z\to\infty$) giving $W_r$ with $r={\overline{N}}$.

Standard power counting applies, producing all counter-terms 
potentially needed 
to eliminate all ultraviolet divergence in perturbation theory. 
With a standard 
gauge fixing method that is independent of $\call$, 
BRST symmetry is preserved
and the counter-terms have to be BRST invariant. 

$\sigma$ has dimension length and therefore the fermions have zero dimension. 
Requiring gauge invariance leaves, in addition to the standard counter-terms, 
$N-1$ new 
dimensionless counter-terms $[\bar\psi\psi]^k$, $k=1,..,N$ with
dimension one coefficients reflecting linear perimeter divergences in the Wilson
loops underlying our observable. Setting $\mu=0$, conjugation invariance is preserved,
acting as follows:
\be
\psi \to\bar\psi,~~ \bar\psi \to \psi,~~ A_\mu \to A_\mu^\ast
\ee
For odd $k$, $[\bar\psi\psi]^k$ changes sign. Therefore, these counter-terms can
be set to zero. The parameter $\mu$ can be turned on perturbatively. By power
counting, this could induce a new ultraviolet divergence, but it is at 
most logarithmic and can be absorbed in $\mu$. 
This leaves $\left [ \frac{N}{2}\right ]$ independent counter-terms, corresponding to
the perimeter divergences in $W_r$ for $r$ given by a $k$-box column;
conjugation symmetry exchanges $k$ with $N-k$ and we require that it be obeyed by the
counter-terms. 

More new counter-terms, of the form 
$\int \bar\psi\partial\psi [\bar\psi\psi]^k$ also are possible. They would 
correspond to new types of logarithmic divergences. 
These counter-terms 
can be made gauge invariant by minimal substitution. (There also is another
set where the derivative acts on $\bar\psi$.) These terms can be ignored, since
they can be eliminated by a field redefinition:
\be
\psi=\chi+\alpha_1 \chi (\bar\psi\chi ) +\alpha_2 \chi (\bar\psi\chi)^2+...+\alpha_{N-1}\chi (\bar\psi\chi)^{N-1}\equiv \chi F (\bar\psi\chi)
\ee
One can change the integration variables $\psi$ to $\chi$ and the Jacobian will
have the form of the counter-terms already taken into account above. The inverse
transformation is:
\be
\chi=\psi-\alpha_1 \psi (\bar\psi\psi) +\beta_2 \psi (\bar\psi\psi)^2+...+\beta_{N-1}\psi (\bar\psi\psi)^{N-1}\equiv \psi 
{\tilde F} (\bar\psi\psi)
\ee
with $\beta_2,..\beta_{N-1}$ determined by the $\alpha_j$. One has then
\be
\bar\psi\partial\psi= \bar\psi\partial [ \chi F(\bar\psi \chi)] =
\bar\psi[\partial \chi] F(\bar\psi \chi)+ \bar\psi\chi [\partial(\bar\psi\chi)] 
F^\prime (\bar\psi\chi)
\ee
The last term is a total derivative, and can be dropped due to the 
boundary conditions.

There is a $\left [ \frac{N}{2}\right ]$ real dimensional space 
in which to choose the
finite parts of the divergent quantities made finite by the counter-terms. 
Once this is done, one can calculate to arbitrary order in perturbation theory.
From the result, by taking derivatives with respect to $z$ at
$z=0$ one gets perturbative expressions for the $N$ coefficients of $z$ in
$P(z,\call)$. The expressions for each coefficient can be re-exponentiated, producing the final perturbative expression for $P(z,\call)$ as a rank $N$ 
polynomial in $z$. 
Alternatively, one could have computed each coefficient 
independently in exponentiated form and never employed the fermions. 

On the $\left [ \frac{N}{2}\right ]$ dimensional submanifold 
of all possible finite part choices 
one has 
enough freedom to ensure that the so obtained 
$P(z,\call)$ will have all its zeros on the unit circle. This is so because
preserving charge conjugation guarantees that the polynomial in $z$ will be
palindromic with real coefficients for any parameter choice~\cite{palindromic}. 
Outside this submanifold 
$P(z,\call)$ will have some complex roots for small enough $\area$. 

Comparing with the known results from diagrammatic analysis, the fermionic representation
is seen to produce correct free parameter counting in this case.

\subsection{A large $N$ phase transition}

Using a lattice regularization one can define $P(z,\call)$ even beyond
perturbation theory. One can make sure that
this fully defined $P(z,\call)$ has all its zeros on the unit circle for any $\area$ 
~\cite{three-d,ourjhep}.

There is numerical evidence for the following: $P(z,\call)$ exhibits a nonanalytic
behavior as the loop $\call$ is kept at fixed shape $\hat\call$ and scaled.
We first take the large $N$ limit in the 't Hooft prescription at 
fixed $z$.  
The nonanalyticity occurs first at $z=1$ as the loop is scaled down from
a large size. 
Wilson loops in $SU(N)$ gauge theory in two, three and four dimensions all 
exhibit this infinite $N$ phase transition as they are shrunk from a 
large size to a small one; in the course of this scaling down, the support of the eigenvalue distribution of the untraced Wilson loop unitary matrix contracts from encompassing the entire unit circle, 
to a small arc centered at $z=-1$ on the unit circle.
An analogous effect takes place in the two dimensional 
principal chiral model for $SU(N)$~\cite{pcm}.

At finite ultraviolet cutoff, before the addition of the multi-fermion 
counter-terms, we observe that the large $N$ transition occurs when the
the spectrum of $D_1 (\call)\equiv \partial_\sigma - i\vpot(\sigma)$, 
which was gap-less for large $\area$, 
opens a gap as $\area$ decreases through $\area_c$. 
The spectrum of $\Omega(\call)$ directly determines the spectrum of $D_1(\call)$, 
which can be viewed as a Dirac operator in one Euclidean dimension~\cite{oddoverlap}.
This observation will partially motivate 
the generalization in the next section.

The universality class of this transition is 
that of a random multiplicative ensemble of unitary matrices.
The transition was discovered by Durhuus and Olesen~\cite{duol} when 
they solved the Makeenko-Migdal~\cite{makeenko} loop equations in two dimensional planar QCD. 
The nature of the transition is now better understood. The authors of ~\cite{blaizot} 
proposed a relation to Burgers' 
turbulence~\cite{burgers} and there are exact results in 2D
supporting their view ~\cite{myburgers}.

For finite $N$, in two dimensions, $\Omega(\call)$ 
is an $SU(N)$ matrix diffusing on the $SU(N)$ group manifold.
This holds approximatively also in four dimensions: see 
~\cite{thies}. The
``diffusion'' is imagined at fixed $\hat\call$ 
and $\area$ plays the role of diffusion time.
At infinite $N$ the transition occurs at $\area=\area_c$. 
For $N\gg 1$, $z\sim 1$ and at $\area\sim\area_c$, $P(z,\call)$ 
[with $\hat\call$ fixed] is described by a universal function, with one, or perhaps two,
non-universal, $\hat\call$ dependent constants. One constant sets the scale
of $\area-\area_c$ and the other sets the scale of $N$ itself. 
There is a possibility that the latter is actually always equal to unity on account 
of the discrete nature of $N$, even though in the $\frac{1}{N}$ expansion 
this does not come in.  

For $N\gg 1$ there is a sharp crossover in the critical region. This 
crossover separates small loops of shape ${\hat \call}$ from large ones with 
identical shape. The crossover connects two distinct regimes. In the perturbative
regime, one has, up to terms depending on the choices made for the finite
parts
\be
P(z,\call) -(1+z)^N {\mathop {\sim}_{\area\to 0}} \;\frac {h_0({\hat\call},N,z)}{\log [\area \Lambda^2_{SU(N)} ]},
\ee
while in the nonperturbative regime we have
\be
P(z,\call) -(1+z^N) {\mathop {\sim}_{\area\to \infty}} (z+z^{N-1}) 
h_\infty ({\hat\call}, N) e^{-\sigma_1 \area}\ee

If we had an effective string algorithm 
to compute in the nonperturbative regime and
managed to carry out a perturbative 
calculation for small loops the two would be
connected at $N\gg 1$ by a known universal function and, by matched asymptotics, we
would obtain the ratio $\frac{\sigma_1}{\Lambda^2_{SU(N)}}$. 
A renormalization scheme that effectively 
fattens the Wilson loop introduces~\cite{ourjhep} 
an additional scale which is finite in units of $\Lambda_{SU(N)}$ and which would
alter in a calculable way the asymptotic behavior as $\area\to 0$. 

To ensure that $P(z,\call)$ has all its zeros on the unit circle for
any area, a necessary condition is
\be
\frac{1}{d_r} W_r (\call ) <1 
\ee
for all $r$ ($d_r$ is the dimension of $r$) and all $\area > 0$ at fixed ${\hat \call}$.
However, many reasonable renormalization 
prescriptions would violate the above inequality for
small enough $\area$.

\section{The new observable}

\subsection{Fermionic representation and ultraviolet divergences}

One would prefer an observable with no linear divergences and less free parameters after renormalization than $P(z,\call)$. 

As mentioned already, the Feynman diagrams themselves are those
of QCD. We decide to restrict now the fermion lines to a two dimensional
embedded closed smooth manifold, $\Sigma$, 
and add the required Dirac indices. We ensure 
that gauge invariance is preserved. To keep matters simple, we want the
two dimensional manifold to be flat in the induced metric from $R^4$. 
We take $\Sigma$ to be a 
torus of sides $l_{1,2}$. The boundary conditions on the fermions
are chosen as antiperiodic in both directions. Now the fermions will have
dimension 1/2 and renormalization will not require terms with more than 4 fermions.
Thus, the number of counter-terms will no longer grow with $N$. Moreover,
if we keep the fermions massless there will be a continuous chiral symmetry
eliminating the last potential linear divergence. This will hold also when the fermion
mass is reinstated. 

Formally, if we take $l_2\to 0$ we get two copies of
the fermion system used for $P(z,\call)$. Thus, the new 
observable ought to still carry
the essential information carried by the old one. 

Consider now a uniform scaling of the 
embedded torus: $l_{1,2}\to \rho l_{1,2}$ and
introduce a shape parameter $0< \tau=\frac{l_1}{l_2}< \infty $ 
which is left invariant by this
scaling. At fixed $\tau$, as $\rho$ is varied, 
maximally separated fermions will feel confining forces
for $\rho \gg 1$ and $\frac{1}{\rm distance}$ 
forces for $\rho \ll 1$. We guess therefore that
the two regimes will be separated by a crossover 
which will become a phase
transition at infinite $N$. As $\rho$ decreases 
through some critical value, the
spontaneously broken 
chiral symmetry will be restored. 
To some extent, this is a reincarnation
of an old idea in~\cite{inst}, 
which in turn was motivated by~\cite{casher}.

At finite cutoff, before the
addition of the 4 fermion counter-terms, 
this large $N$ transition will be reflected in the
eigenvalue spectrum of $D_2(\Sigma)$, 
the Dirac operator acting on the 
torus fermions. This is analogous to an 
observation we made in the previous section.
However, it is well known that on the broken side the portion
of the spectrum of $D_2(\Sigma)$ that 
is close to zero is described by a simple
random matrix model~\cite{chiralrmt}. 
This fact has become a potent tool for numerically
determining that chiral symmetry is spontaneously broken. 
The random matrix description 
ceases to hold when chiral symmetry is restored. 

Here are some equations summarizing the above. 
The embedded torus $\Sigma$ is defined by $x(\sigma)$, $\sigma$ 
is short for $\sigma_\alpha$, 
with $\alpha=1,2$.\symbolfootnote[1]{From the context it should be clear when $\sigma_{1,2}$ refers to these coordinates and when to the first two
string tensions among the $\sigma_k$.}
\be
x_1(\sigma)=\frac{l_1}{2\pi}\cos\frac{2\pi \sigma_1}{l_1};
~~x_2(\sigma)=\frac{l_1}{2\pi}\sin\frac{2\pi \sigma_1}{l_1};
~~x_3(\sigma)=\frac{l_2}{2\pi}\cos\frac{2\pi \sigma_2}{l_2};
~~x_4(\sigma)=\frac{l_2}{2\pi}\sin\frac{2\pi \sigma_2}{l_2}
\ee
The induced metric on the torus is
\be
d\sigma^2=d\sigma_1^2 + d\sigma_2^2
\ee
We define a two component gauge potential $\vpot_\alpha$ on the torus by
\be
\vpot_\alpha=A_\mu (x(\sigma))
\frac{\partial x_\mu}{\partial \sigma_\alpha}
\ee
The Dirac matrices on the torus are
\be
\gamma_1=\pmatrix{0 & 1\cr1 & 0}~~\gamma_2=\pmatrix{0 & -i\cr i & 0}
\ee

The new observable is:
\be
Q(\mu,\Sigma)=\frac{
\int [dA_\mu ] [d\bar\psi d\psi ]  
e^{-\frac{1}{4g^2}\int d^4 x \tr F^2 +\int_\Sigma d^2\sigma
\bar\psi(\sigma)[\gamma_\alpha \partial_{\sigma_\alpha} - \mu - i\gamma_\alpha\vpot_\alpha(\sigma)]\psi(\sigma)}}{
\int [dA_\mu ] e^{-\frac{1}{4g^2}\int d^4 x \tr F^2}}
\ee
The two dimensional massless Dirac operator is:
\be
D_2 (\Sigma )=\gamma_\alpha \partial_{\sigma_\alpha} -  i\gamma_\alpha\vpot_\alpha(\sigma)
\label{diracb}
\ee

Denoting the Hermitian generators of $su(N)$ in the fundamental representation by
$T^j$, $j=1,..,N-1$ the nonabelian fermion current coupled to $\vpot$ is given by
\be
J^j_\alpha (\sigma ) = \bar\psi (\sigma) \gamma_\alpha T^j \psi (\sigma)
\ee
The abelian vector current is
\be
J_\alpha = \bar\psi (\sigma) \gamma_\alpha \psi (\sigma)
\ee
Power counting and symmetries allow for two local four-fermion counter-terms at $\mu=0$:
\be
\LL_1 = J^j_\alpha J^j_\alpha,~~~\LL_2= J_\alpha J_\alpha
\ee
$\LL_1$ can be replaced by a chiral invariant 
linear combination of terms made out
of the product of two terms bilinear in the fermions, each a 
$SU(N)$ singlet. Suppose one integrates out the Yang Mills and ghost 
fields first. Now, one keeps the four fermion terms of order $g^2$, but
drops all higher order terms in $g^2$. We leave aside the question how
valid this approximation is, but intuitively it should capture correctly
some of the physics if $\Sigma$ is small in units of $\Lambda_{SU(N)}^{-1}$. 
This makes it
possible to use standard methods~\cite{grossneveu} to solve
the fermion model exactly in the large $N$-limit, 
even though the induced four fermion term is
non-local~\cite{njlqcd}. The required $\LL_1$ counter-term can also be
exactly taken into account in the large $N$ limit.
As is well known, this provides expressions for some observables
that are non-perturbative in the 't Hooft coupling $g^2 N$.

Both counter-terms are dimensionless, so there are no ultraviolet divergences worse than
logarithmic. The Thirring term  
$\LL_2$ is probably not strictly needed, and would
add a free parameter to the theory if included~\cite{niemi}. 
$\LL_1$ is certainly needed, as indicated above. 
To identify the ultraviolet divergences it is enough to 
consider the case where our closed $\Sigma$ is replaced by
an infinite two dimensional plane. The $A_\mu$ 
Feynman gauge propagator in four dimensional Fourier
space induces an effective $a_\alpha$ propagator in two dimensions:
\be
\int_{p_\|^2 + p_\perp^2 \le \Lambda^2} \frac{d^2 p_\| d^2p_\perp}{(2\pi)^4} \frac{f(p_\|^2)}{p_\|^2 + p_\perp ^2 }=
\int_{p_\|^2 \le \Lambda^2} \frac{d^2 p_\|}{(2\pi)^3} f(p_\|^2) \log[\Lambda/|p_\||]
\label{logdiv}
\ee

We have not yet carried out a detailed direct 
diagrammatic analysis of all ultraviolet divergences.

\subsection{General properties}

The new observable thus achieves the main objectives of reducing all ultraviolet
divergences to logarithmic and making the number of 
needed counter-terms $N$-independent. The new observable is more 
amenable to perturbation theory, as infinite sets of diagrams can be summed
using results from large $N$ vector models and non-analytic dependencies
in the Yang Mills coupling can be generated~\cite{njlqcd}. 
A calculation of the relevant $\beta$-functions for this system is left for the future.
Perturbation theory should work for small $l_{1,2}$, but for 
large $l_{1,2}$ we need something different. Possibly, an effective 
two dimensional field theory could be found and then we would need to
match it to the four dimensional underlying YM theory.
Alternatively, although we do not
have a clear candidate for an effective string description, we could guess one.
If we have an effective description of
the Wilson operators for large loops by 
using open strings whose ends are 
restricted to the curve $\call$, maybe a 
similar effective description employing open strings now restricted 
to end on $\Sigma$ 
(a sort of D-brane) would work for large $\Sigma$. 

One issue left to address is how confinement is reflected by the new observable, 
qualitatively and quantitatively. We already noted that the limit $\tau\to 0$ takes
the new observable into the old one, so confinement will be seen. 
There are several new ways in which the new observable would reflect confinement.
It is well known~\cite{makeenko} that one can write a formal expression for
the chiral condensate in terms of a sum over Wilson loops. Here, the
chiral condensate would be a 2D one, but the Wilson loops, although restricted
to $\Sigma$ will have four dimensional values\symbolfootnote[2]{
We are assuming $N\gg 1$ 
where the feedback of the fermions can be neglected.}
. Large loops will exhibit the area law
because of four dimensional confinement and those loops would make
a contribution to the condensate which is similar to the contribution made in
two dimensional gauge theory if the 
two fundamental string tensions are matched. 
Note however that the minimal area 
may not lie on $\Sigma$ and therefore even for
large loops one cannot expect an asymptotically perfect match. In addition,
in two dimensional gauge theory there is no shape dependence, while in four dimensions
there is. Nevertheless, 
from the point of view of the fermionic condensate the
four dimensional character of the ambient YM theory reflects itself mostly at 
small Wilson loops. It is unclear by how much the relation between 
the $\sigma_1$ and the condensate would differ in our system from an
effective two dimensional one.
Note that two dimensional Yang Mills 
theory with the inclusion of $\LL_{1,2}$ 
is still exactly
solvable at infinite $N$~\cite{njlymtwod}.

The finite nature of the torus reflects itself in the free fermion propagator.
If the gauge fields were truly two dimensional, at infinite $N$ the dependence
on the size of the torus would disappear from intensive quantities~\cite{contred}.
The mechanism behind this relies on the extra finite-temperature-like $Z^2(N)$
symmetry of the gauge sector 
which acts non-trivially on Polyakov loop operators on the torus.
These symmetries are absent when the gauge fields are induced from the four 
dimensional world. However, for a large enough torus the four dimensional 
Wilson loop operators interacting with the fermions as Polyakov loop operators
have eigenvalue distributions that are very flat. The deviation from
total flatness is exponentially small in the area because of four dimensional
confinement. 
Thus, up to a small correction, the fermions free energy per unit torus
area will become independent of the volume at very large $N$. More precisely, 
\be 
\lim_{N\to\infty}\frac{1}{Nl_1 l_2}\log [Q(\mu,\Sigma) ]
\ee 
would approach its infinite $l_{1,2}$ limit (at fixed $\tau\ne 0,\infty$) 
with a correction that goes as $e^{-\sigma_1 l_1 l_2}$.
We see that in the large $N$ limit the fundamental string tension controls
the large scale behavior of the new observable, just as that of the old one.

A last issue of comparison is that in the old observable the fermions can 
be easily integrated out (this is how they were introduced in the first place)
leaving one with explicit expressions in terms of Wilson loop operators.
The fermions of the new observable can also be integrated out, leaving a more complicated observable behind. If $\Sigma$ were a two sphere the answer
has an explicit form~\cite{polywieg}. For a torus some care is required 
with loops wrapping the torus and we are not sure whether a closed expression can be written down. Recall that a local form of the answer requires the introduction
of a third dimension along which the gauge field is smoothly deformed. In our application one needs not invent such a deformation: $\Sigma$ can be ``filled'' 
in two topologically distinct ways inside the $R^4$ and many possible 
deformations are available. 

\subsection{Hamiltonian version}

One can imagine a Hamiltonian formulation, in which we choose $l_2=\infty$.
Then the infinitely heavy finitely separated 
source and sink picture associated with an ordinary rectangular Wilson loop,
and used for the extraction of the heavy quark-antiquark potential, has
a simple generalization. Instead of a nailed down pair of fundamentals
we have a circle on which any number of fundamentals can travel and many
types of states will contribute. The locations of the 
charges are still restricted in four
dimensions, but not to two separated points, but rather to a common circle. So, in particular, 
as time evolves, we guess that states made out 
of diametrically opposite quark-anti-quark pairs, rotating around each
other, would develop. This is closer to the semi-classical picture we usually invoke
when extracting the experimental string tension from meson Regge 
trajectories in QCD than the static heavy quark potential
we get from traditional Wilson loops. 
Moreover, since there is no fundamental matter in the bulk, the rotating
string along the diameter cannot just decay by pair formation, as it 
would in QCD~\cite{cnn}. 
To be sure, while 
our observable lets the pair members rotate
round each other, it still constrains the radius of their trajectory to a fixed number, 
so the four dimensional argument 
cannot be fully carried over. 
In short, the new observable provides an opportunity to look at mesonic 
states of definite angular momentum in some direction. 

Replacing the finite torus $\Sigma$ by an infinite cylinder does not eliminate the
large $N$ phase transition.  The 
Hamiltonian system is expected to undergo 
a large $N$ phase transition of spontaneous chiral symmetry breaking as the compact
direction is shrunk, in
analogy with the finite temperature Gross-Neveu model~\cite{ulli}.

\section{Numerical results for large $N$}

\subsection{General observations}

One cannot put a smooth closed loop on a hypercubic lattice. One has to allow
corners, and there are extra logarithmic divergences associated with these. 
When dealing with Wilson loops the definition of the observable we used 
in our numerical work~\cite{ourjhep,three-d,pcm} also eliminated these divergences. 

Similarly, one cannot put a smoothly embedded torus on the lattice. However, 
the worst new ultraviolet divergences induced by the 
extra folds one requires are weaker 
than the leading ones, which now are just logarithmic. 
Thus we can use a torus embedded into a four dimensional hypercubic lattice
in which the circle of circumference $l_1$ in $\Sigma$ is replaced by a square 
of perimeter $L_1$ (in lattice units and divisible by 4) and similarly
the circle of circumference $l_2$ in $\Sigma$ is replaced by a square 
of perimeter $L_2$. It is now straightforward to adapt the overlap~\cite{overlap}
action to the lattice $\Sigma$ so constructed. This allows us to have exact 
chiral symmetry on the lattice. 

Since the fermions only contribute at order $N$ to the free energy, to determine
what happens at $N=\infty$ it suffices to keep them quenched, that is their
feedback on the distribution of the gauge fields can be ignored~\cite{chissb}. This simplifies
the numerical work by a significant amount. 

Equation~(\ref{logdiv}) tells us that in order to go to the continuum limit
we need to add a 4 fermion term on the embedded surface. 
Its sign will be the 
opposite of the standard sign; after a Fierz transformation and a Hubbard-Stratonovich 
decoupling, the kernel of the quadratic fermion action no longer obeys
a property under hermitian conjugation that simplifies the numerical computation of its
eigenvalues. Our numerical tests below did not include 4 fermion terms. We looked
for numerical evidence for spontaneous chiral symmetry 
breaking at infinite $N$ in the 
lattice regularized theory. More work would be needed to
go to the continuum limit. 

Even as a pure lattice study, 
our numerical work is just exploratory; our main objective was to find
examples of parameters for which we can be reasonably sure that
chiral symmetry is spontaneously broken at $N=\infty$ 
and other examples where
we can be reasonably sure that chiral symmetry is preserved by the vacuum of the
fermionic system even at $N=\infty$. This provides evidence for the existence of
at least one large $N$ phase transition. The simplest assumption is that
there exists only one such transition, but we have certainly not ruled out 
numerically the existence of more transitions than one. 
Our extrapolations to infinite $N$ are simplistic and would need to be
refined in future work. 

\subsection{Parameter ranges}

The four dimensional gauge field at fixed lattice coupling,
$b=\frac{1}{g^2N}$ and given $N$, is generated in the standard manner
on lattices of sizes $6^4$ and $10^4$. We looked at the
range of $b$-values $0.348$--$0.367$. The highest $b$ value is determined
by the requirement that the $10^4$ torus still be in the confining phase at
$N=\infty$. The lowest $b$ value is determined by the requirement that
we stay out of the infinite $N$ strong coupling phase, at least by metastability. 
The smaller volume can be used for smaller $b$-s in the range.

A two dimensional finite flat torus of size $4L\times 4L$
is constructed by the Cartesian product of a closed $L\times L$ loop in the (1,2) plane 
by another closed $L\times L$ loop in the (3,4) plane. We looked at sizes
$4L=8,16$.

A ``semi-infinite'' two dimensional torus of size $4L$ in its finite
direction is embedded in the four dimensional lattice 
forming one closed $L\times L$ loop in the (1,2) plane
and another loop in the $3$ (or $4$) direction  closed by winding around
the lattice. At $N=\infty$, because of continuum reduction, 
the torus may be viewed as infinite in the winding direction
so long as we are in the confined phase. Again, we looked at $4L=8,16$.

We construct the two dimensional overlap operator with anti-periodic
boundary conditions on the two dimensional torus, using the 
four dimensional gauge fields residing on the links of the embedded 
torus. We compute the two lowest positive eigenvalues of the overlap Dirac operator for
$100$ random distinct translations of the two dimensional torus on the
four dimensional lattice. In this manner we get an average value
for the two lowest eigenvalues, $\bar\lambda_i$, $i=1,2$. 
We also obtain the distribution of $r=\frac{\lambda_1}{\lambda_2}$, denoted by $P(r)$,
for each four dimensional gauge field configuration. 
We obtain an estimate for $\langle \bar\lambda_i\rangle$, $i=1,2$, 
$\langle \bar r\rangle$ and $\langle P(r)\rangle $ by averaging
over several configurations.

\subsection{Finite torus}

\subsubsection{Broken chirality}

We present one example in the phase of broken chirality ($L=16$,  
$b=0.350$ on a $10^4$ lattice). 
Figures \ref{fig1} and \ref{fig3} show data for $N=13$ and $N=19$ 
and provide evidence for the absence of a spectral gap 
(Figure \ref{fig1}) and for chiral 
random matrix behavior of the eigenvalue ratio distribution (Figure \ref{fig3}).
The two lowest
eigenvalues in Figure \ref{fig1} extrapolate at $N=\infty$ to small
negative numbers which we interpret as saying that both eigenvalues
go to zero.
One can also calculate the average eigenvalue ratio and extrapolate it to $N=\infty$;
one obtains a number comfortably close to the chiral random matrix theory prediction,  
$0.37674227...$. 

The absence of a gap and the agreement with chiral random matrix theory are strong
indicators that chiral symmetry is spontaneously broken in the $N\to\infty$ limit at
this choice of the parameters $L$ and $b$.

\FIGURE{
\centering
\includegraphics[scale=0.4]{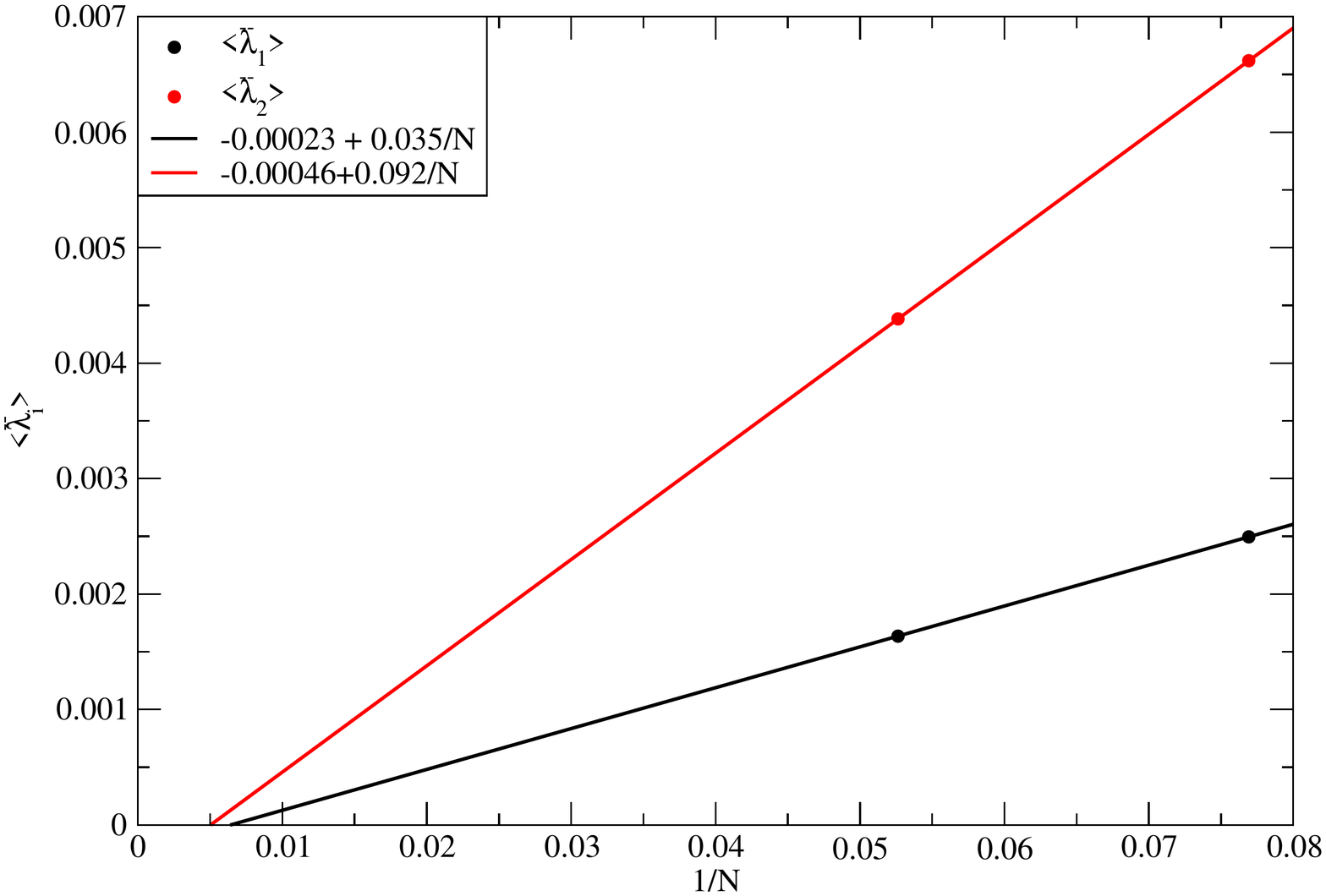}
\caption{\label{fig1} Lowest two eigenvalues as a function of $1/N$
at $b=0.350$ for a 
$16\times 16$ torus embedded in a $10^4$ lattice.}}

\FIGURE{
\centering
\includegraphics[scale=0.4]{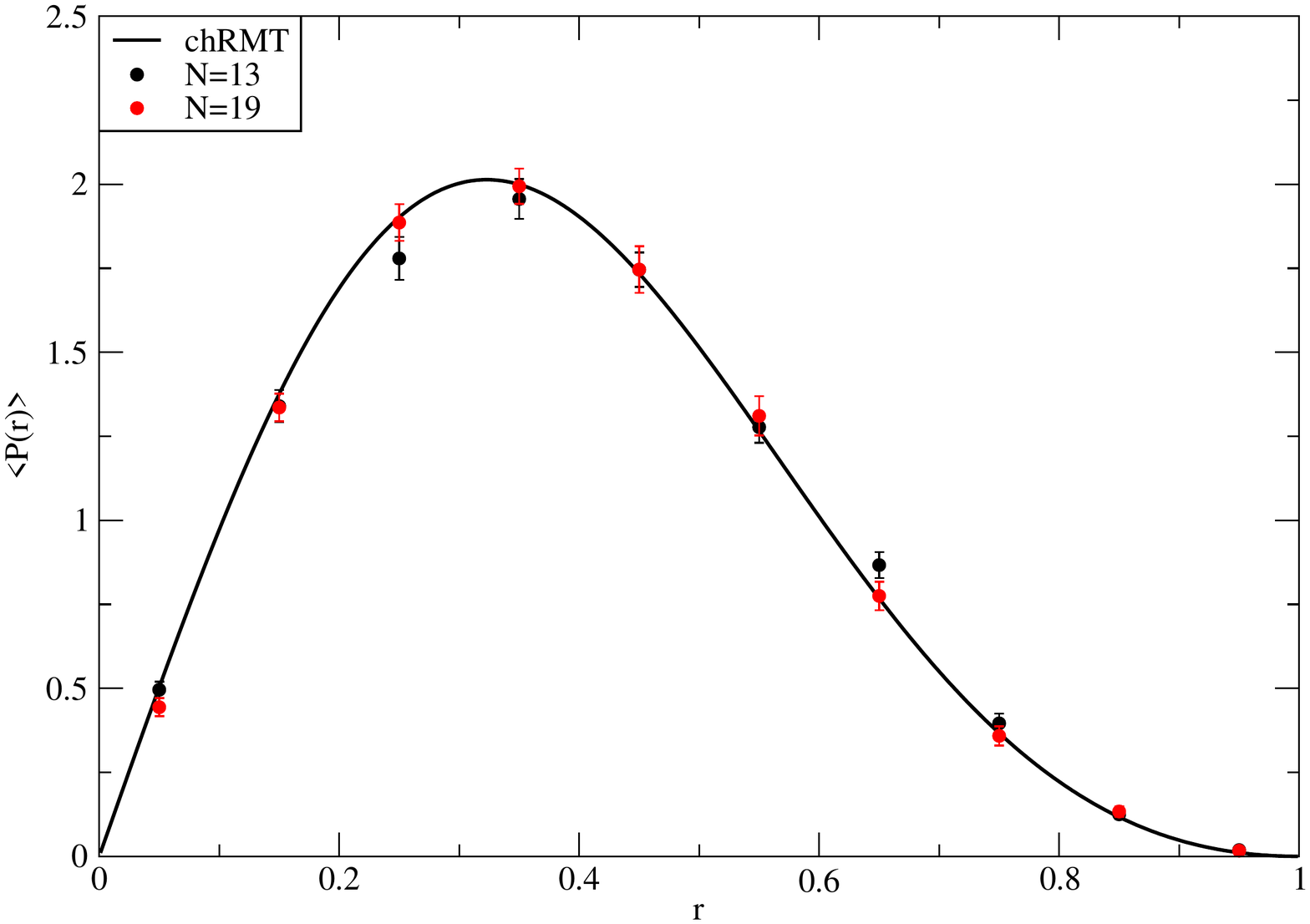}
\caption{\label{fig3}Distribution of the ratio of the two lowest eigenvalues
for $N=13$ and $N=19$ at $b=0.350$ for a 
$16\times 16$ torus embedded in a $10^4$ lattice.}}

\subsubsection{Preserved chirality}

Our other example ($L=8$, $b=0.348$ on a $6^4$ lattice) is 
in the phase of preserved chirality.
We consider a larger range of $N$ values, 
$N=13,19,23,31,37$ because this case requires a more thorough 
investigation of the $N$-dependence: chiral random matrix behavior
might set in at only high values of $N$, and we want to rule this out. 
 Not only is the limit of the smaller
eigenvalue significantly different from zero, but the separation
of the second eigenvalue from the first also has a nonzero limit.
Figure \ref{fig9} shows that the distribution of
the ratio of the two lowest eigenvalues gets more peaked closer
to unity as $N$ is increased, becoming more and more different from 
the chiral random matrix theory result. 

The existence of a gap separating the lowest eigenvalue from
zero at infinite $N$ implies that chiral symmetry is preserved at this set
of $L=8$, $b=0.348$ values. 

\FIGURE{
\centering
\includegraphics[scale=0.4]{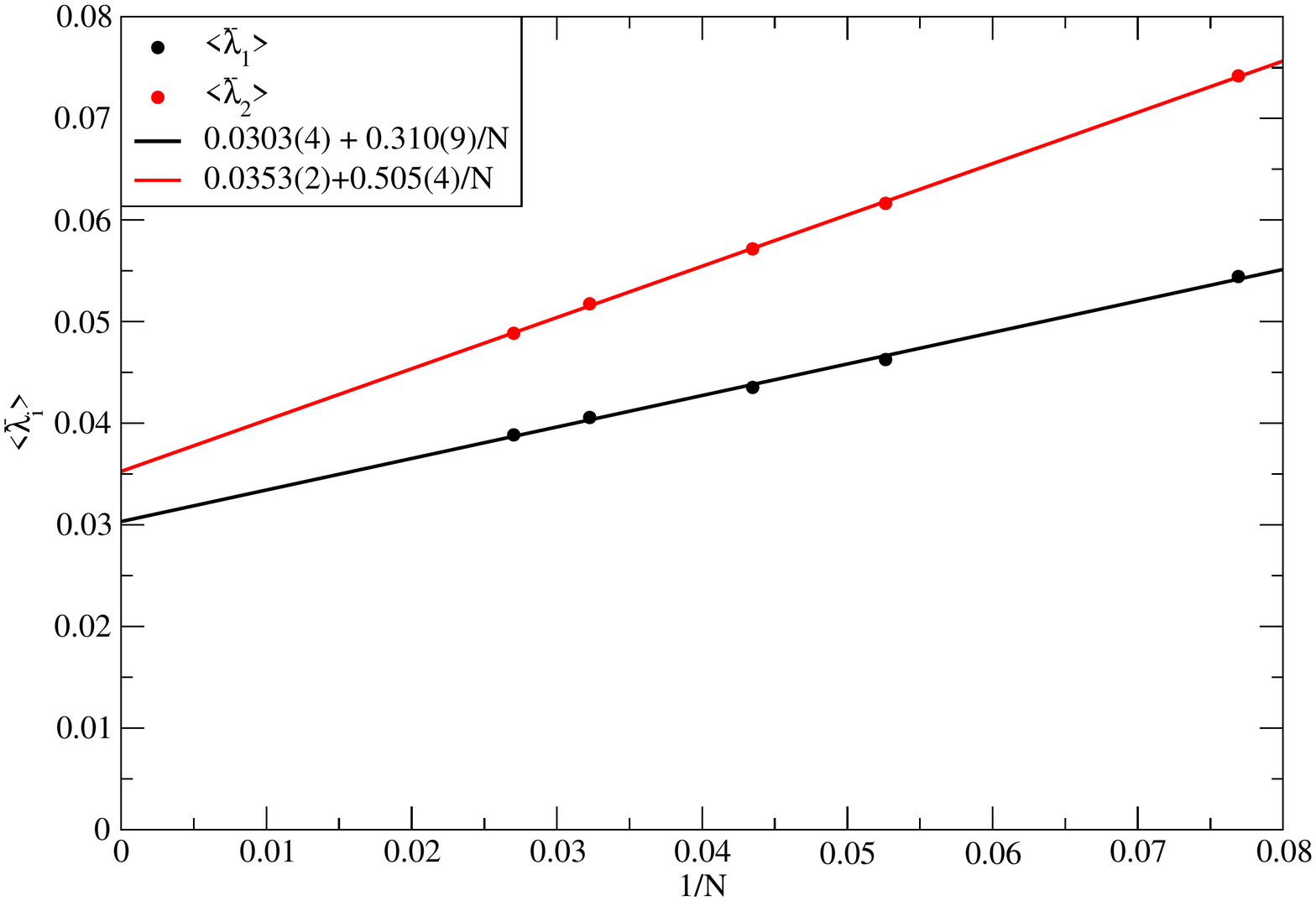}
\caption{\label{fig7}Lowest two eigenvalues as a function of $1/N$
at $b=0.348$ for a 
$8\times 8$ torus embedded in a $6^4$ lattice.}}

\FIGURE{
\centering
\includegraphics[scale=0.4]{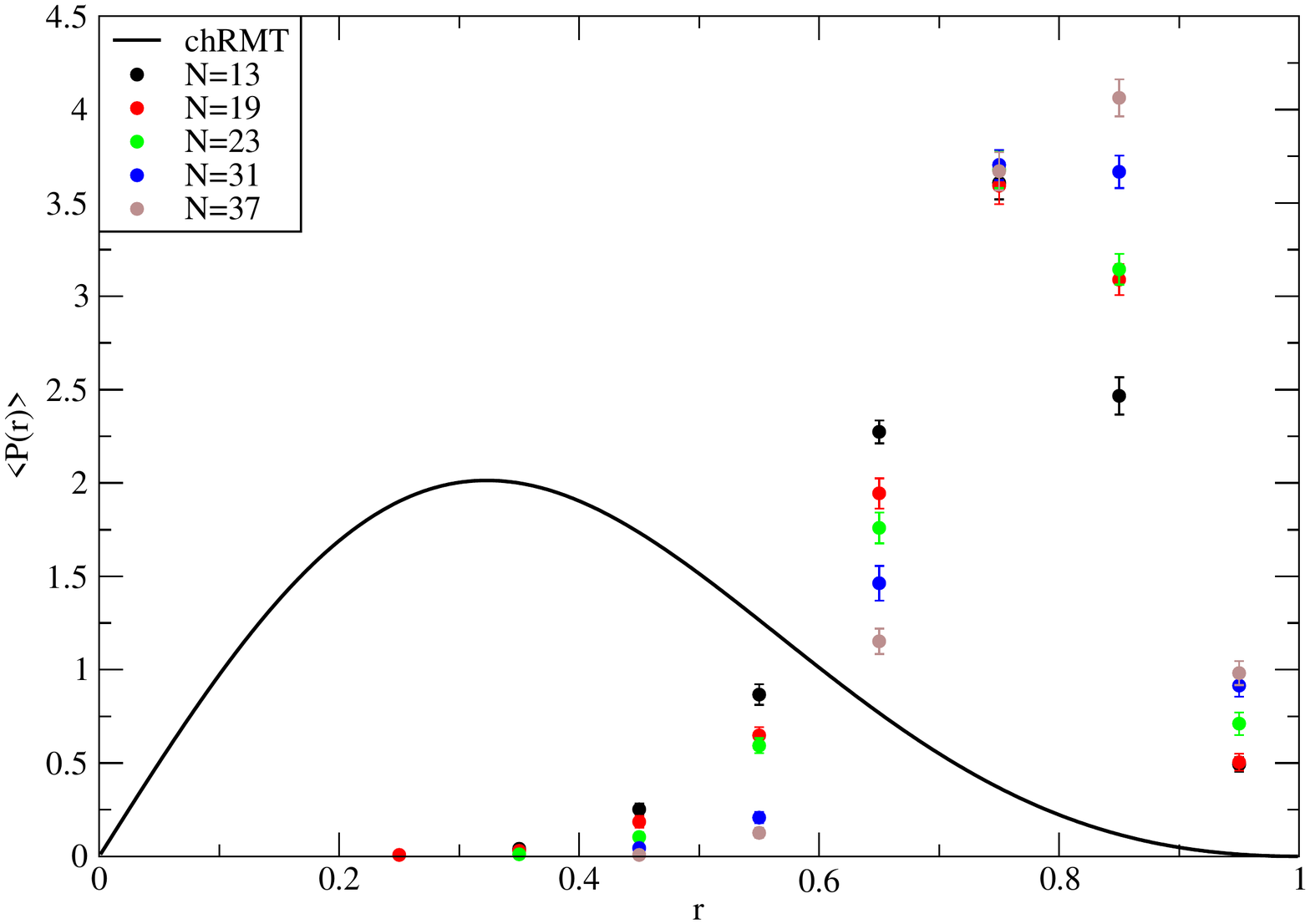}
\caption{\label{fig9}Distribution of the ratio of the two lowest eigenvalues
for $N=13,19,23,31,37$ at $b=0.348$ for a 
$8\times 8$ torus embedded in a $6^4$ lattice.}}

\subsection{Torus with one infinite extent}

The behavior of the lowest two eigenvalues on a torus 
with one infinite extent follows the same pattern as on 
the finite torus.

\subsubsection{Broken chirality}

We consider a $16\times 10$ torus on a $10^4$ lattice and set 
$b=0.363$. The $N$ values we used are $N=13,19,28,27$. 
The fits to the two lowest 
eigenvalues in Figure \ref{fig10} are again interpreted 
as saying that both eigenvalues extrapolate to zero at $N=\infty$.

Figure \ref{fig12} shows that the distribution of 
the ratio of the two lowest eigenvalues agrees 
with the prediction
from chiral random matrix theory when extrapolated to infinite $N$. 
The average ratio again approaches the chiral random matrix theory number. 

We conclude that chiral symmetry is spontaneously broken at $N=\infty$ at
this point in parameter space. 

\FIGURE{
\centering
\includegraphics[scale=0.4]{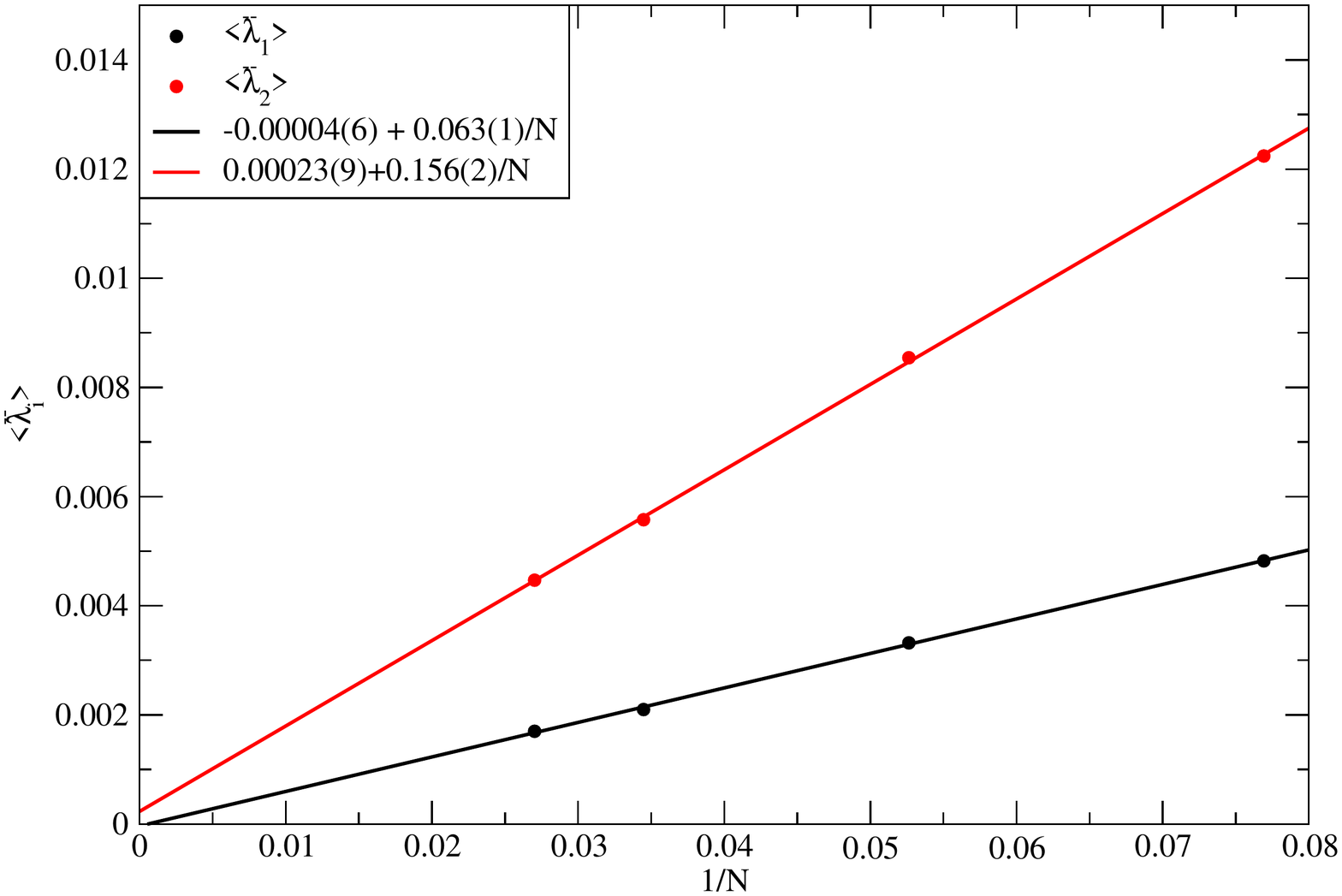}
\caption{\label{fig10}Lowest two eigenvalues as a function of $1/N$
at $b=0.363$ for a 
$16\times 10$ torus embedded in a $10^4$ lattice.}}

\FIGURE{
\centering
\includegraphics[scale=0.4]{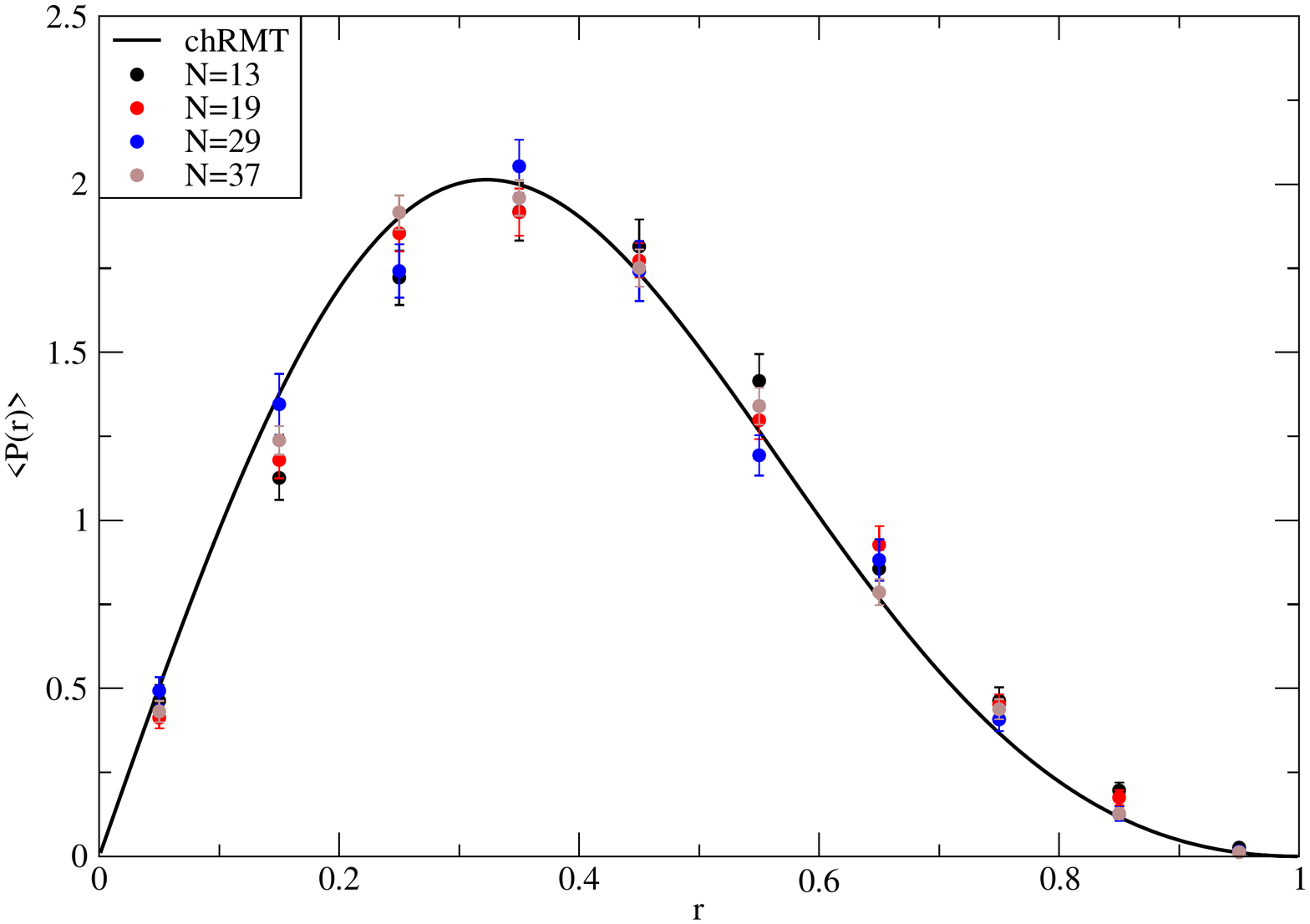}
\caption{\label{fig12}Distribution of the ratio of the two lowest eigenvalues
for $N=13,19,29,37$ at $b=0.363$ for a 
$16\times 10$ torus embedded in a $10^4$ lattice.}}

\subsubsection{Preserved chirality}

We now take an $8\times 10$ torus on a $10^4$ lattice at 
$b=0.363$ for $N=29$ and $N=37$. The plot of the two lowest
eigenvalues as a function of $1/N$ in Figure \ref{fig13} indicates 
the existence of a  gap at $N=\infty$.
The two lowest
eigenvalues in Figure \ref{fig13} extrapolate to values which show
the presence of a gap at $N=\infty$. 
Figure \ref{fig15} shows that the distribution of
the ratio of the two lowest eigenvalues gets more peaked closer
to unity as $N$ is increased, departing from the prediction of chiral
random matrix theory. 

We conclude that chiral symmetry is preserved also at $N=\infty$ in this example.

\FIGURE{
\centering
\includegraphics[scale=0.4]{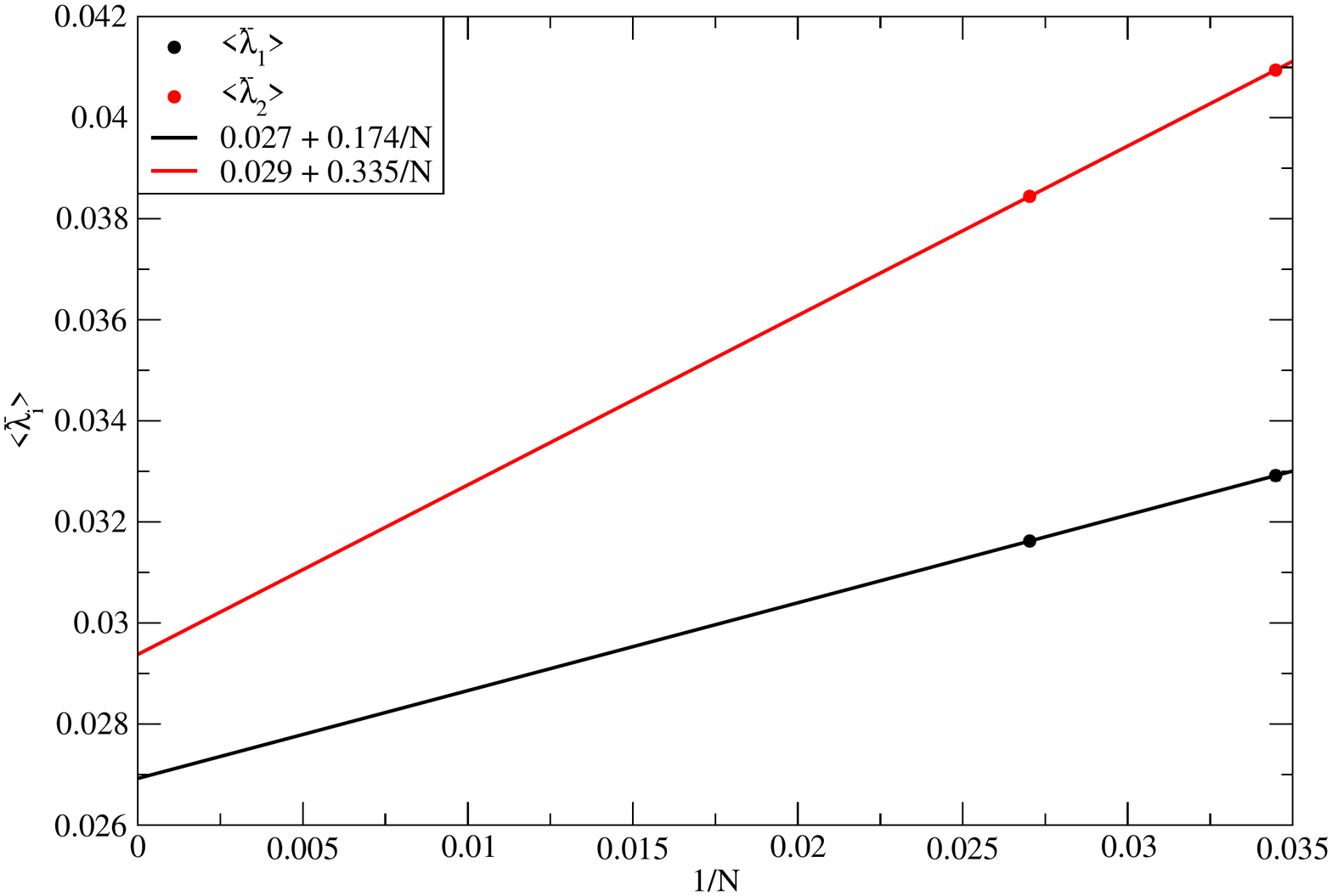}
\caption{\label{fig13}Lowest two eigenvalues as a function of $1/N$
at $b=0.363$ for a 
$4\times 10$ torus embedded in a $10^4$ lattice.}}

\FIGURE{
\centering
\includegraphics[scale=0.4]{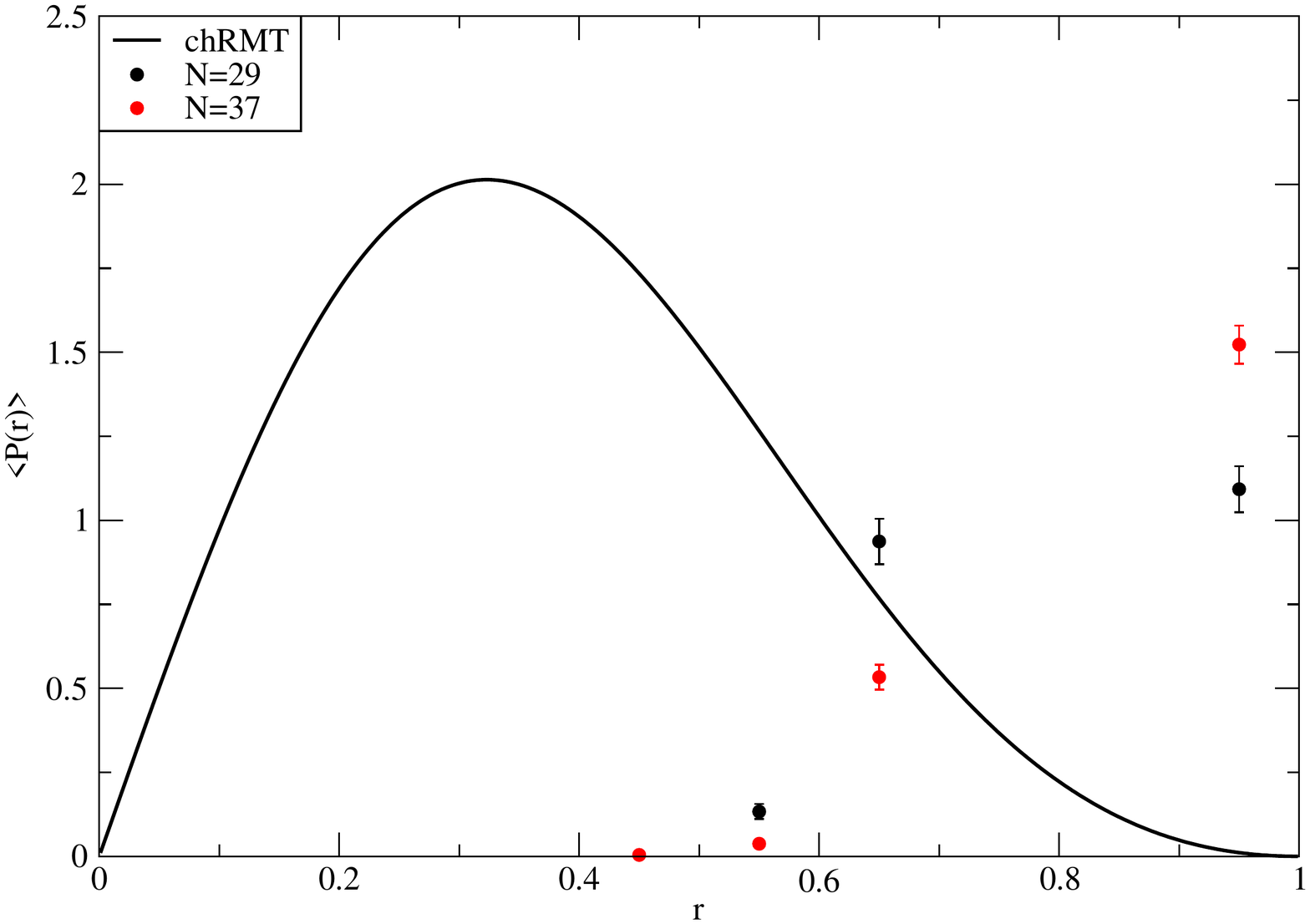}
\caption{\label{fig15}Distribution of the ratio of the two lowest eigenvalues
for $N=29$ and $N=37$ at $b=0.363$ for a 
$4\times 10$ torus embedded in a $10^4$ lattice.}}

\subsection{Comments on numerical work}

We have carried out more types of fits in the above selected examples and
also simulations at other parameters sets. These studies showed
that the simplistic extrapolations of $\frac{1}{N}$ to zero have systematic
errors that can overcome the statistical ones. We also found points in
parameter space where we could not credibly make a determination about
the status of chiral symmetry at infinite $N$. Therefore we cannot rule
out the existence of some more esoteric phases in the large $N$ limit. 
If more esoteric phases do turn out to exist, there will be more than one 
large $N$ phase transitions and the large $N$ structure of the crossover
may be more complicated than we would like. 

These issues need to be addressed. It is possible that the
computational resources at our disposal at the moment won't be able to
reach the values of $N$ needed to settle these questions conclusively.

\section{Summary and Discussion}

Our main point was to introduce a system coupling two dimensional 
fermions to four dimensional gauge fields. 
This provides new non-local observables in pure Yang Mills theory and 
we gave reasons why they are interesting. We 
hope that these observables in pure gauge theory will produce parameters expressible in terms of $\Lambda_{SU(N)}$ by an 
asymptotic matching of perturbation theory to a yet unknown, 
effective, systematic, description of long distance physics. 
We provided numerical evidence that these new observables 
vary with scale through a 
crossover where a nonanalytic behavior will set in at
infinite $N$. We hypothesize that at $N=\infty$ 
the crossover becomes a phase 
transition governed by a well understood
universality class. 

If this hope pans out, we could match parameters between the two descriptions
and use them in other circumstances. For example, one could consider fermions
living on the infinite $z,t$ plane while the gauge fields it couples to
live on infinite Minkowski space $x,y,z,t$. It is now convenient to pick
the light 
cone gauge in the $z,t$ directions and consider scattering events of objects made out of fermions in the $z,t$ plane. Unlike in the cylinder 
or torus case, for a scattering event characterized by a single scale, one
expects that even at infinite $N$ 
the crossover as that scale is varied from short to long will be 
smooth. The infinite $N$ phase transitions we are 
studying are observable dependent and so are their universality classes.
The basic idea is then to use an observable which has a large $N$
phase transition in order to relate the parameters of a long distance
effective theory to those convenient 
at short distances, and then exploit the 
general applicability 
of the long distance theory to calculate 
other observables, which are smooth
in scale even at $N=\infty$.

The infinite spacetime could be taken to be three dimensional, in which case
the ultraviolet divergence~(\ref{logdiv}) goes away. It is possible that in this
case one can do without a four fermion counter-term in the continuum. 
One cannot embed a torus flatly
in the ambient three dimensional spacetime, 
and therefore one might as well use a sphere and deal with its curvature.
On the lattice the sphere would be replaced by the surface of a cube 
equipped with 8 singular corners.
We leave further study of this case for the future.

\subsection*{Acknowledgments.}

R.N. acknowledges partial support by the NSF under grant number PHY-0854744.
HN acknowledges partial support by the DOE under grant
number DE-FG02-01ER41165.
HN notes with regret that his research has for a long time been 
deliberately obstructed by his high energy colleagues at Rutgers.   
HN wishes to thank A. Schwimmer for many conversations. HN acknowledges
a private conversation held at Zakopane in May 2009, in which M. Nowak expressed his
intuition that the large $N$ phase transition in Wilson loops 
reminds one of spontaneous chiral symmetry breaking. 
HN also thanks R. Lohmayer and T. Wettig for conversations and L. Stodolsky for
several electronic mail exchanges.

\end{document}